\documentclass[aps,prl,twocolumn,showpacs,superscriptaddress,letterpaper,10pt]{revtex4-1}

\usepackage[colorlinks=true,citecolor=blue]{hyperref}
\usepackage{mathrsfs,graphicx,amssymb}
\usepackage[fleqn]{amsmath}
\usepackage{calrsfs}

\newcommand{\ket}[1]{\left| #1 \right\rangle}
\newcommand{\bra}[1]{\left\langle #1\right |}

\DeclareMathOperator{\Tr}{Tr}

\begin{document}

\title{Dipole-Induced Electromagnetic Transparency}

\author{Raiju Puthumpally-Joseph}
\affiliation{Universit\'e Paris-Sud, Institut des Sciences Mol\'eculaires d'Orsay (CNRS), F-91405 Orsay, France}

\author{Maxim Sukharev}
\affiliation{Science and Mathematics Faculty, School of Letters and Sciences, Arizona State University, Mesa, Arizona 85212, USA}

\author{Osman Atabek}
\affiliation{Universit\'e Paris-Sud, Institut des Sciences Mol\'eculaires d'Orsay (CNRS), F-91405 Orsay, France}

\author{Eric Charron}
\affiliation{Universit\'e Paris-Sud, Institut des Sciences Mol\'eculaires d'Orsay (CNRS), F-91405 Orsay, France}

\date{\today}

\begin{abstract}
We determine the optical response of a thin and dense layer of interacting quantum emitters. We show that in such a dense system, the Lorentz redshift and the associated interaction broadening can be used to control the transmission and reflection spectra. In the presence of overlapping resonances, a Dipole-Induced Electromagnetic Transparency (DIET) regime, similar to Electromagnetically Induced Transparency (EIT), may be achieved. DIET relies on destructive interference between the electromagnetic waves emitted by quantum emitters. Carefully tuning material parameters allows to achieve narrow transmission windows in otherwise completely opaque media. We analyze in details this coherent and collective effect using a generalized Lorentz model and show how it can be controlled. Several potential applications of the phenomenon, such as slow light, are proposed.
\end{abstract}

\maketitle

Light-matter interaction has been a topic of intense research for many decades. It is currently experiencing a significant growth in the area of nano-optics \cite{Novotny.book}. Light scattering by a system of nanometric size is an example where important applications can be foreseen. The theoretical description of light scattering is very well understood when dealing with individual quantum emitters such as atoms or molecules \cite{Cohen.Wiley.1992}. In the case of two (or more) strongly interacting emitters, and in general for large densities, the physics is far more complex since the behavior of the ensemble of emitters cannot be described anymore as the sum of their individual response. In this case, the field experienced by an emitter depends not only on the incident field but also on the one radiated by all its neighbors. The latter are also affected by the emitter, thus leading to a very complex highly coupled dynamics which must be described self-consistently.

For an oscillating dipole of resonant wavelength $\lambda_0$, high densities $n_0$ are achieved when \mbox{$n_0\,\lambda_0^3 \geqslant 1$}, \textit{i.e.} when there is more than one emitter in the volume associated with the dipole wavelength \cite{JPB.44.195006}. In this situation, strong dipole-dipole couplings come into play, and collective excitation modes quickly dominate the optical response of the sample. This results usually in an enhancement of light-matter interaction. This cooperative effect is clearly observed in the superradiance or superfluorescence processes initially discussed by Dicke \cite{PR.93.99}.

The topic of strong dipole-dipole interactions has recently been the subject of a considerable interest in the context of quantum information with cold atoms \cite{RMP.82.2313}, following an early proposal by Jaksch \textit{et al} to use dipole blockade as a source of quantum entanglement \cite{PRL.85.2208}. This initial proposal, limited to two interacting dipoles, was soon extended to many-atom ensemble qubits \cite{PRL.87.037901}. With highly excited Rydberg atoms, this regime can be achieved for atomic densities as low as $10^{10}\,\mathrm{cm}^{-3}$ \cite{PRL.105.193603,Science.336.887}. Higher densities, of the order of $10^{15}\,\mathrm{cm}^{-3}$, are typically required for ground state atoms. As an example of cooperative effects, collective Lamb and Lorentz shifts were recently measured in a thin thermal atomic vapor layer similar to the system studied here \cite{PRL.108.173601}.

Following these last developments, the present paper deals with a theoretical study of the optical response of a thin dense vapor of quantum emitters, atoms or molecules. We show that, in such systems, strong dipole-dipole interactions can be used to manipulate the spectral properties of the light scattered by the sample. We also show that in the presence of overlapping resonances \cite{JCP.137.094302}, the medium may become partially transparent for a particular frequency which can be controlled to a certain extent. In addition, the radiation at the neighboring frequencies is nearly perfectly reflected, opening the way to potential applications in optics.

\begin{figure}[!t]
\includegraphics[width=0.99\columnwidth]{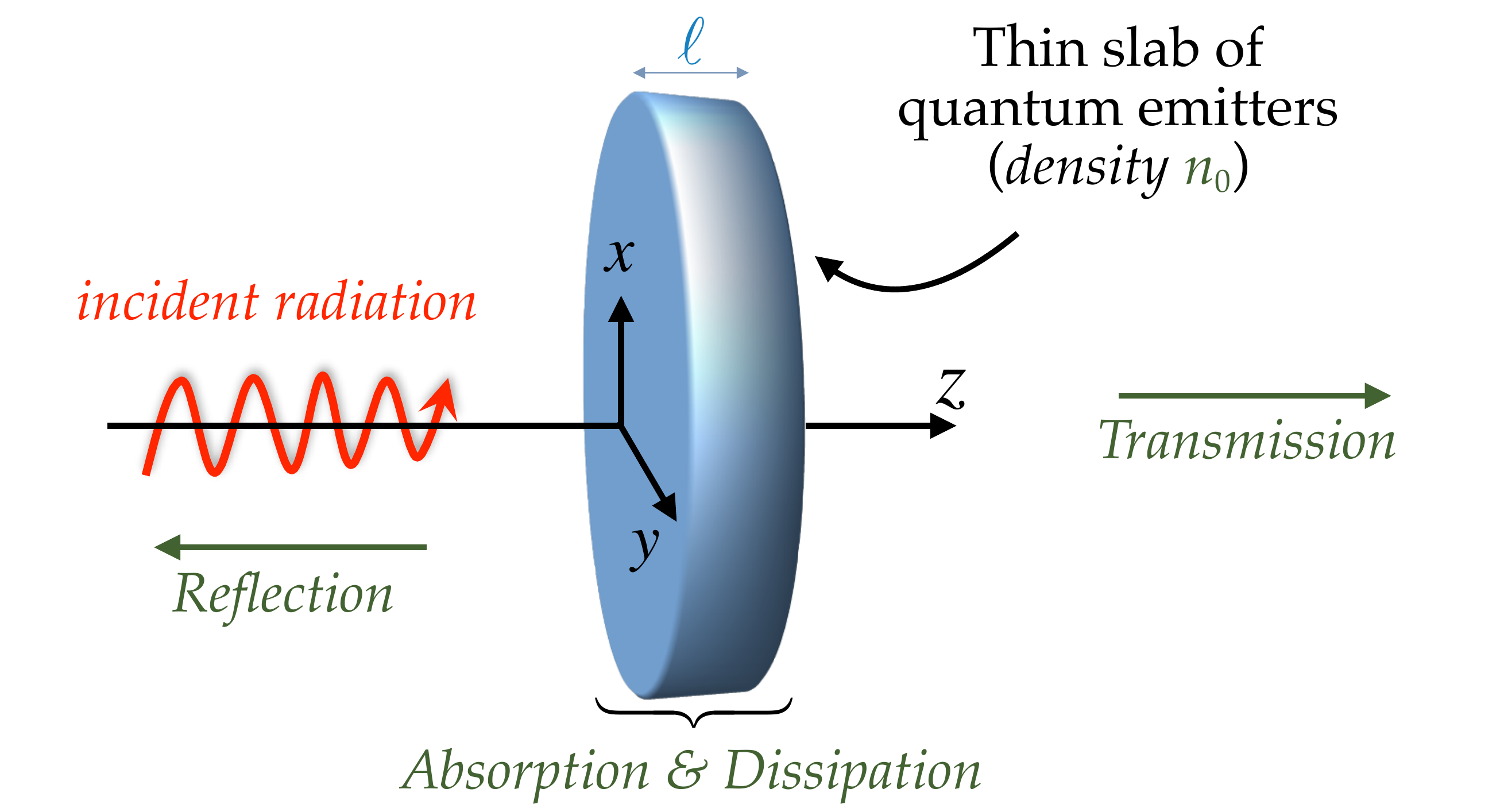}
\caption{\label{fig1}(Color online) Schematic view of the thin dense vapor of quantum emitters interacting with the incident field.}
\end{figure}

As shown in Fig. \ref{fig1}, we consider a thin layer of quantum emitters whose transverse dimension ($z$-axis) is denoted by $\ell$. The longitudinal dimensions of the slab in the $x$ and $y$ dimensions are assumed to be much larger than $\ell$. The dipoles considered here are two-level emitters, with states labeled as $\ket{0}$ and $\ket{1}$ . Their associated energies are \mbox{$\hbar\omega_0$} and \mbox{$\hbar\omega_1$}. $\omega_{01} = \omega_{1} - \omega_{0}$ denotes the Bohr frequency. The density matrix $\hat\rho(z,t)$ \cite{Blum.PP.1996,PRA.84.043802} describing the quantum dynamics satisfies the dissipative Liouville-von Neumann equation
\begin{equation}
 \label{eq:Liouville}
 i\hbar\partial_t\hat\rho = [ \hat H , \hat \rho ] -i\hbar \hat\Gamma \hat\rho,
\end{equation}
where $\hat H = \hat H_0 + \hat V(z,t)$ is the total Hamiltonian and $\hat\Gamma$ is a superoperator taken in the Lindblad form \cite{Breuer.OUP.2002}, describing relaxation and dephasing processes under Markov approximation. The field free Hamiltonian reads
\begin{equation}
 \label{eq:H0}
 \hat H_0 = \hbar\omega_0 \ket{0}\!\!\bra{0} + \hbar\omega_1 \ket{1}\!\!\bra{1},
\end{equation}
and the interaction of the two-level system with the electromagnetic radiation is taken in the form
\begin{equation}
 \label{eq:Vint}
 \hat V(z,t) = \hbar\Omega(z,t)\,\big( \ket{1}\!\!\bra{0} + \ket{0}\!\!\bra{1} \big)
\end{equation}
where $\Omega(z,t)$ is the local instantaneous Rabi frequency associated with the transition dipole $\mu_{01}$. In Eq.\,(\ref{eq:Liouville}), the non-diagonal elements of the operator $\hat\Gamma$ include a pure dephasing rate $\gamma^*$, and the diagonal elements of this operator consist of the radiationless decay rate $\Gamma$ of the excited state. The total decoherence rate is denoted by $\gamma = \gamma^* + \Gamma/2$. Equations (\ref{eq:Liouville})-(\ref{eq:Vint}) lead to the well-known Bloch optical equations \cite{Allen.Wiley.1975,JCP.138.024108} describing the quantum dynamics of a coupled two-level system. It is assumed that the system is initially in the ground state $\ket{0}$.

The incident radiation is normal to the slab and propagates in the positive $z$-direction (see Fig.\,\ref{fig1}). It is represented by a transverse electric mode with respect to the propagation axis and is characterized by one in-plane electric and one out-of-plane magnetic field components, namely $E_x(z,t)$ and $H_y(z,t)$. Time-domain Maxwell's equations in such a geometry read
 \begin{eqnarray}
  \mu_0\,\partial_t H_y       & = & -\partial_z E_x\\ \label{Faraday}
  \epsilon_0\,\partial_t E_x & = & -\partial_z H_y -\partial_t P_x \label{Ampere}
 \end{eqnarray}
The system of Maxwell's equations is solved using a generalized finite-difference time-domain technique where both the electric and magnetic fields are propagated in discretized time and space \cite{PRA.84.043802,Taflove.AH.2005}. The macroscopic polarization \mbox{$P_x(z,t)= n_0\langle \hat{\mu}_{01} \rangle=n_0\Tr[\hat\rho(z,t)\hat{\mu}_{01}]$} is taken as the product of the atomic density $n_0$ with the expectation value of the transition dipole moment operator $\hat\mu_{01}$. The coupled Liouville-Maxwell equations are integrated numerically in a self-consistent manner. The coupling between Eqs.(\ref{eq:Liouville}) and (\ref{Ampere}) is through the polarization current $\partial_t P_x$ due to the quantum system taken as a source term in Ampere's law (\ref{Ampere}) but, as discussed below, this is not sufficient in the case of high densities.

An exact treatment of light scattering in the presence of strong interactions between a large number of quantum emitters is extremely difficult. It has been shown that an efficient and accurate approach consists in the introduction of a local field correction to the averaged macroscopic electric field $E_x(z,t)$ \cite{Lorentz.Dover.1952}. In this mean-field approach, the individual quantum emitters are driven by the corrected local field
\begin{equation}
 \label{eq:Eloc}
E_{\text{local}}(z,t) = E_x(z,t) + P_x(z,t)/(3\epsilon_0).
\end{equation}
This local field $E_{\text{local}}(z,t)$ enters the dissipative Liouville equation (\ref{eq:Liouville}) through the Rabi frequency $\Omega(z,t)=\mu_{01}\,E_{\text{local}}(z,t)/\hbar$. It is well known that the replacement of $E_x(z,t)$ with $E_{\text{local}}(z,t)$ leads to a frequency shift in the linear response functions of the medium for large densities, the so-called Lorentz-Lorenz (LL) shift $\Delta = n_0 \mu_{01}^2 / (9\hbar\epsilon_0)$ \cite{Lorentz.Dover.1952,Jackson.Wiley.1975}.

\begin{figure}[!t]
\includegraphics[width=0.99\columnwidth]{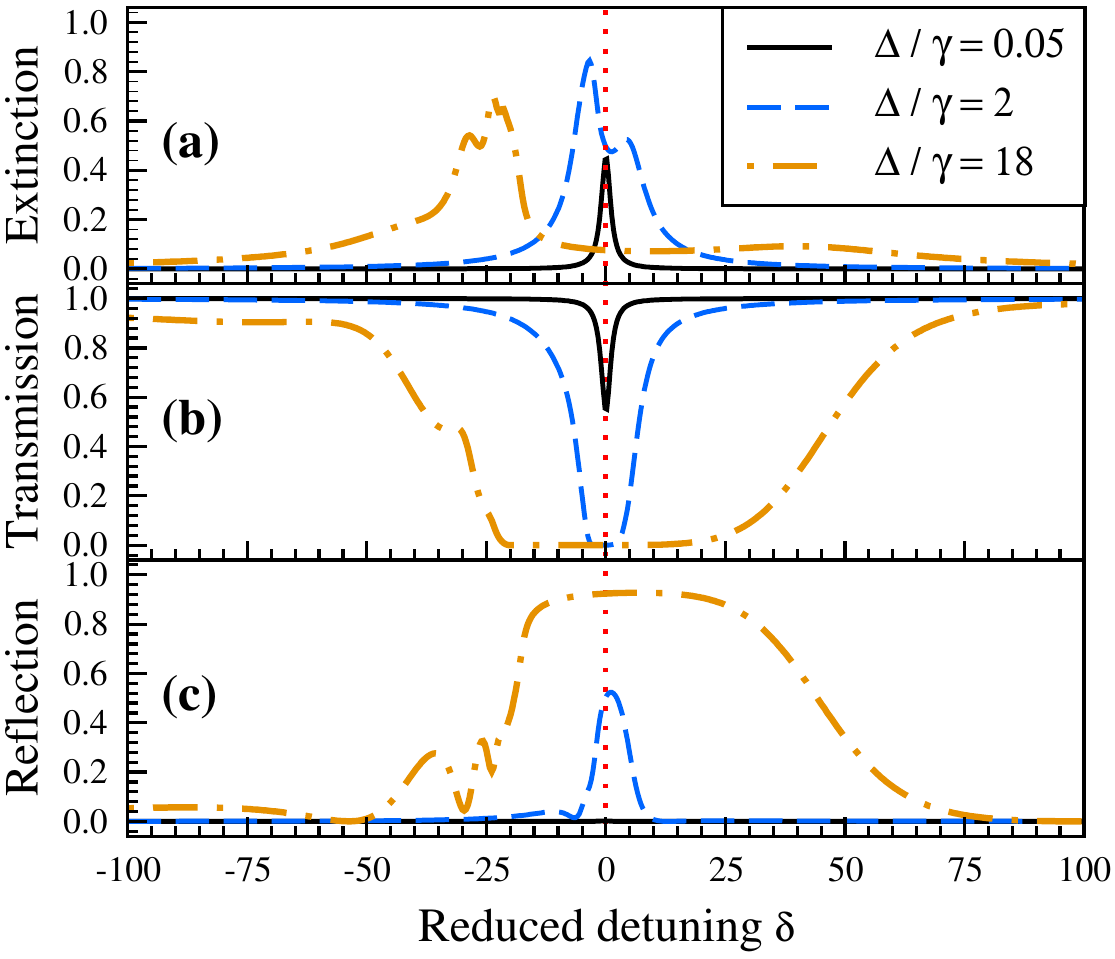}
\caption{\label{fig2}(Color online) Extinction (a), transmission (b) and reflection (c) probabilities as a function of the reduced detuning for a thickness $\ell = \lambda_0/1.55 = 400$\,nm. The decay and pure dephasing rates are $10^{11}$ and $10^{12}$\,Hz. The solid black lines correspond to a weak density $[\Delta / \gamma = 0.05]$, the dashed blue lines are for an average density $[\Delta / \gamma = 2]$, and the dash-dotted brown lines stand for a large density $[\Delta / \gamma = 18]$.}
\end{figure}

Figure \ref{fig2}  shows the calculated one-photon extinction, transmission and reflection spectra as a function of the reduced detuning $\delta=(\omega-\omega_{01})/\gamma$ at three different densities. These spectra are obtained via the computation of the normalized Poynting vector on the input and output sides of the layer \cite{PRA.84.043802,JCP.138.024108}. It is important to note that for weak densities (solid black lines) the extinction spectrum [panel (a)] shows a typical Lorentzian lineshape of half-width $\gamma$. Light absorption affects the transmission spectrum [panel (b)] such that a hole is observed, and no reflection is seen in panel (c). An increased density (blue dashed lines) leads to a splitting of the extinction signal into two lines: the red-shifted line corresponds to a configuration where the dipoles oscillate in-phase with the incident field, whereas the blue-shifted line corresponds to an anti-parallel configuration where the induced dipoles oscillate out-of-phase with this field. In addition, the hole seen in the transmission spectrum broadens significantly and looses its Lorentzian shape. Concurrently, a strong reflection signal shows up at the transition frequency. The optical response of the medium changes dramatically at high densities (dash-dotted brown lines). The medium is then characterized by a collective dipole excitation which cancels out transmission over a very large window around the transition frequency. In this frequency range, almost total reflection is observed. This collective effect can be understood from the extended Lorentz model we introduce below.

\begin{figure}[!t]
\includegraphics[width=0.99\columnwidth]{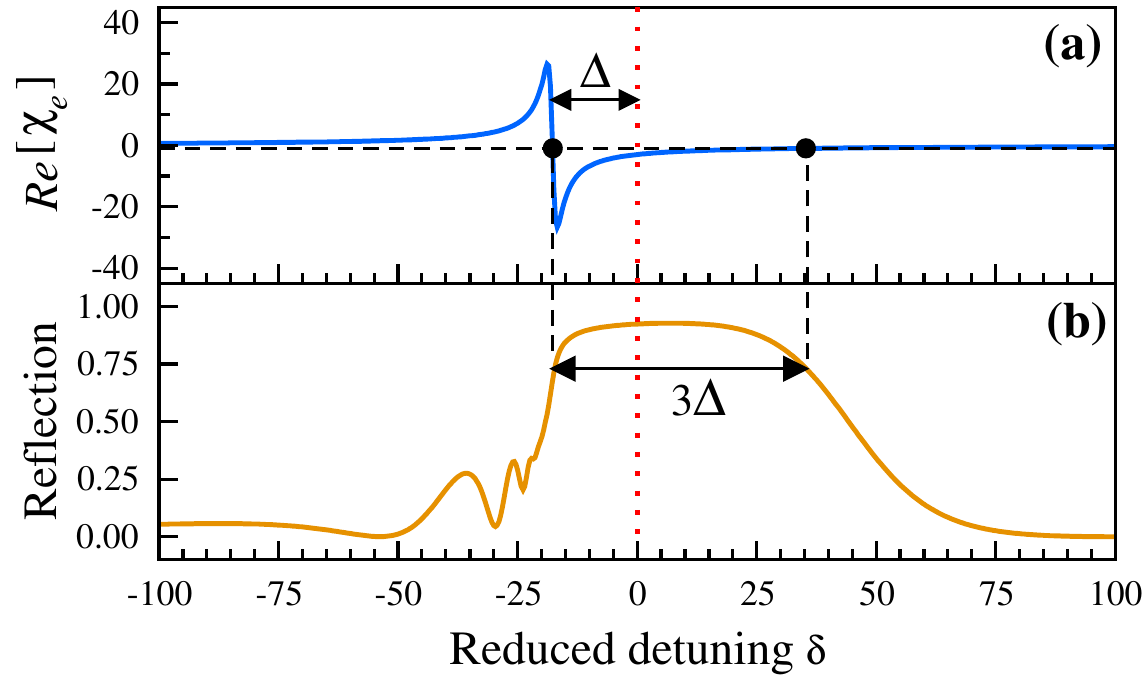}
\caption{\label{fig3}(Color online) (a) Real part of the electric susceptibility $\chi_e$ calculated from our extended Lorentz model as a function of the reduced detuning, for $\Delta / \gamma = 18$ (large density). The reflection probability is shown in panel (b). The two dots in panel (a) indicate the frequencies at which $Re[\chi_e] = -1$.}
\end{figure}

In this model, the dipoles, driven by the electric field, experience linear restoring and classical damping forces. The time evolution of the macroscopic polarization can then be written as \cite{Lorentz.Dover.1952}
\begin{equation}
 \label{eq:Pol}
 \partial_{tt}P_x + \gamma\,\partial_{t}P_x + \omega_{01}^2\,P_x = \epsilon_0 \omega_p^2 [E_x + P_x/(3\epsilon_0)]
\end{equation}
where $\omega_p$ denotes the plasma frequency. Compared to the usual formulation of the classical Lorentz model, we have added here the local field correction $P_x/(3\epsilon_0)$. In addition, with the assumption that the maximum amplitude of oscillation of the dipoles in the absence of a driving field is given by the quantum harmonic oscillator length, we obtain the plasma frequency $\omega_p = \sqrt{6\,\omega_{01}\Delta}$, where $\Delta$ is the LL shift. Finally, in the particular case of a monochromatic excitation, Eq.\,(\ref{eq:Pol}) is easily solved, and the electric susceptibility $\chi_e = P_x / (\epsilon_0 E_x)$ is obtained as $\chi_e(\omega) = 6\,\omega_{01}/f(\omega)$, where $f(\omega)=(\omega_{01}^2-2\omega_{01}\Delta-\omega^2 + i \gamma \omega)/\Delta$. Compared to the standard Lorentz model, we observe here a frequency shift $\omega_{01}^2 \rightarrow \omega_{01}^2-2\omega_{01}\Delta$. When \mbox{$\Delta/\omega_{01} \ll 1$}, we see that the resonance frequency $\omega_{01}$ is simply red-shifted by the LL shift $-\Delta$, as expected \cite{Jackson.Wiley.1975}. We see clearly in panel (a) of Fig. \ref{fig3} such a strong redshift of the resonance. The broad reflection window seen in panel (b) for large densities, which was already predicted by Glauber \emph{et al} \cite{PRA.61.063815}, can now be explained using simple considerations. Assuming a non-absorbing medium, and therefore $\gamma=0$, the reflectance $R(\omega)$ at the interface is given by \mbox{$R = | (1 - n) / (1 + n) |^2$}, where $n(\omega) = Re[\sqrt{1+\chi_e(\omega)}]$ is the real part of the refractive index of the slab. We see that total reflection is obtained when $n = 0$ and therefore when $\chi_e(\omega) \leqslant -1$. Our model nicely predicts that this is achieved in the frequency range $[\omega_{01}-\Delta\,,\,\omega_{01}+2\Delta]$. The width of the reflection window is therefore $3\Delta = n_0 \mu_{01}^2 / (3\hbar\epsilon_0)$, as shown in Fig.\,\ref{fig3}. Panels (a) and (b) show that the out-of-phase oscillations of the induced dipoles correspond to a nearly opaque overall sample. Here, the strongly coupled oscillating dipoles emit a radiation which efficiently cancels out the incident field inside the sample, thus leading to high reflection. This phenomenon dominates in case of high densities, where the dipoles coherently cooperate to prevent penetration of the incident radiation in the slab.

\begin{figure}[!t]
\includegraphics[width=0.99\columnwidth]{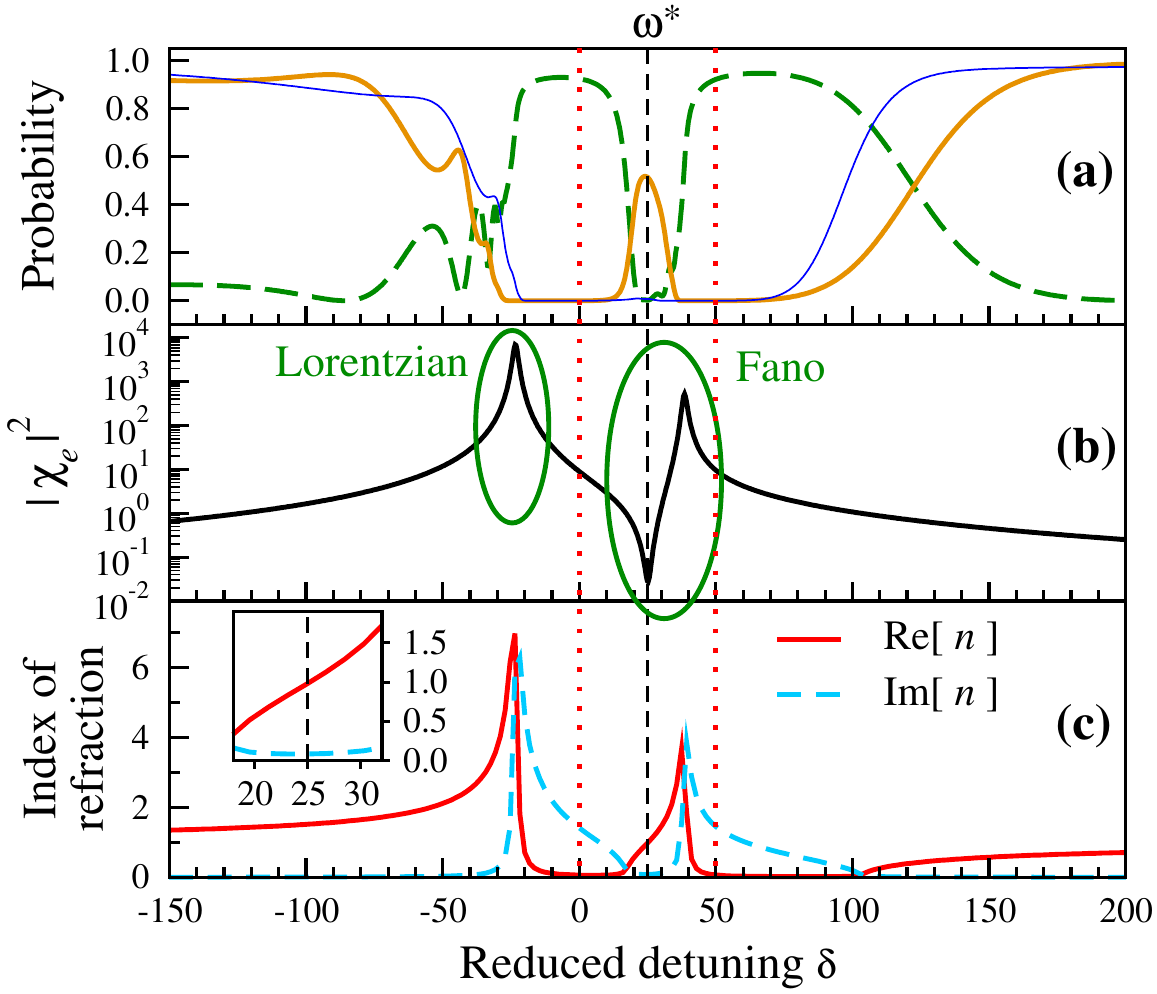}
\caption{\label{fig4}(Color online) Transmission (solid brown line) and reflection (dashed green line) probabilities (a) as a function of the detuning in the case of a dense mixture of two different quantum emitters (see text for details). The thin blue line is the transmission probability when the coupling between the two dipoles is neglected. The electric susceptibility and the refractive index are shown in panels (b) and (c).}
\end{figure}

Let us now use the in-phase \emph{vs} out-of-phase dipoles to manipulate both reflection and transmission. We consider a mixture of two different quantum emitters with densities $n_0$ and $n_0'$ and transition dipoles $\mu_{01}$ and $\mu_{01}'$. Fig.\,\ref{fig4} shows the case of two dipoles with the same decoherence rate $\gamma=\gamma'$ and the same LL shift $\Delta  = 18\gamma = \Delta'$. The two transitions differ by $(\omega_{01}'-\omega_{01})/\gamma = 50$ and the reduced detuning is still defined with respect to the first transition. The two vertical red dotted lines indicate the frequencies of the two transitions. A very peculiar feature shows up in the reflection (dashed green line) and transmission (solid brown line) spectra of Fig.\,\ref{fig4}. At the intermediate detuning $\delta = 25$, a minimum appears in the reflection spectrum. Concurrently, a sharp transmission peak appears at the same frequency. The thin blue line shows the transmission when the coupling between the two dipoles is neglected. Clearly, the strong coupling between the two types of dipole renders the medium transparent in an otherwise opaque region. We note that the position and width of the transparency window are controlled by material parameters, as discussed below.

Let us introduce the coupling between the two dipoles in our extended Lorentz model. The time evolution of the polarization $P$ associated with the first dipole reads
\begin{equation}
 \label{eq:Pol2}
 \partial_{tt}P + \gamma\partial_{t}P + \omega_{01}^2 P = \epsilon_0\omega_p^2 [E_x + (P+P')/(3\epsilon_0)]
 \end{equation}
with an equivalent equation describing the polarization of the second dipole $P'$. The total polarization is then written as $P_x=P+P'$. These two coupled equations can be solved analytically in the case of a monochromatic driving field, yielding
\begin{equation}
 \label{eq:chie2}
 \chi_e(\omega) = \frac{6\omega_{01}'[f(\omega)+2\omega_{01}]+6\omega_{01}[f'(\omega)+2\omega_{01}']}{f(\omega)f'(\omega)-4\omega_{01}\omega_{01}'}
\end{equation}
where $f'(\omega)=(\omega_{01}'^{\,2}-2\omega_{01}'\Delta'-\omega^2 + i \gamma' \omega)/\Delta'$. The square modulus of this electric susceptibility is shown in panel (b) of Fig.\,\ref{fig4}. The resonance observed at the detuning $\delta=-35$ presents the usual Lorentzian profile and marks the frequency at which reflectance reaches a plateau. Another resonance is seen in the region $\delta=20$\,-\,40. This resonance has a Fano profile, characteristic of a quantum interference effect between two indistinguishable excitation pathways \cite{Fano.PR.1961}. Indeed, in the frequency range between $\omega_{01}$ and $\omega_{01}'$, the two transitions overlap and the contributions from the two types of dipoles add up in the macroscopic polarization of the medium. In addition, in this frequency range the two dipoles oscillate in opposite directions (out-of-phase) and one can find a particular frequency $\omega^*$ at which they cancel each other. This comes from the fact that one type of emitters is blue-detuned while the other is red-detuned, leading to opposite signs of their susceptibilities. In the limit $\gamma \ll \Delta$ we obtain
\begin{equation}
 \label{eq:w*}
\omega^{*\,2} = \frac{(\omega_{01}\Delta)\,\omega_{01}'^{\,2} + (\omega_{01}'\Delta')\,\omega_{01}^2}
                                           {(\omega_{01}\Delta)+(\omega_{01}'\Delta')}.
\end{equation}
At this intermediate frequency $\omega^* \simeq \omega_{01} + 25\gamma$ the susceptibility reaches almost zero and the medium becomes transparent. Knowing that $\Delta \propto n_0$ and that $\Delta' \propto n_0'$, it appears that the value of  $\omega^*$ can be controlled in the range $\omega_{01} \leqslant \omega^* \leqslant \omega_{01}'$ by simply adjusting the densities of the two emitters. This transparency phenomenon is reminiscent of EIT \cite{JETP.48.630, PRL.62.1033, PRL.66.2593, Scully.CUP.1997} and we therefore name it ``Dipole-Induced Electromagnetic Transparency'' (DIET). Compared to EIT, the strong coupling induced by the pump laser is replaced by strong dipole-dipole interactions.

\begin{figure}[!t]
\includegraphics[width=0.99\columnwidth]{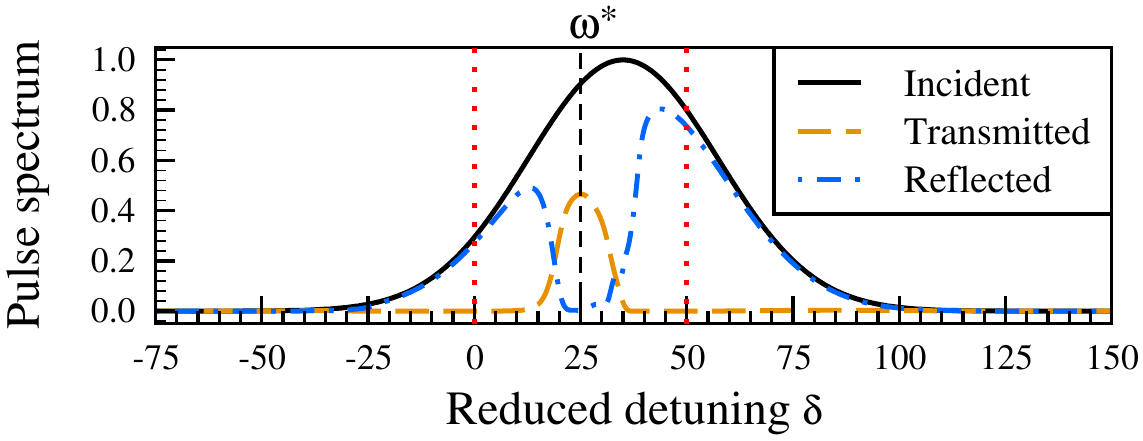}
\caption{\label{fig5}(Color online) Incident, transmitted and reflected pulse spectrum in the case of a dense vapor constituted by a mixture of two different quantum emitters.}
\end{figure}

An example of application is shown Fig.\,\ref{fig5}, where the incident field is a weak 50\,fs pulse of carrier frequency $\omega_{01}+35\gamma$. We show in Fig.\,\ref{fig5} the power spectrum of the incident, transmitted and reflected fields. Clearly, DIET is imprinted in both the reflected and transmitted pulses. Furthermore, as seen in Fig.\,\ref{fig4}(c), at $\omega=\omega^*$ the medium is characterized by a steep dispersion of the real part of the refractive index, whereas its imaginary part, and therefore absorption, is negligible. At this frequency the group velocity  $v_g = c/[n+\omega(dn/d\omega)]$ reaches $v_g \simeq c/250$. We verified numerically that the transmitted pulse is time-delayed by $\ell/v_g$, a proof that DIET can induce slow light \cite{PRL.74.666, PRL.74.2447, Nature.397.594, PRL.82.5229, Science.312.895}. In addition, the slow-down factor can be controlled by changing the material parameters.

In terms of experimental implementations, we believe that DIET could be observed in an atomic vapor confined in a cell whose thickness is of the order of the optical wavelength \cite{PRL.108.173601,PRA.69.065802}. Such systems suffer from inhomogeneous Doppler broadening \cite{PR.89.472}. At room temperature, the induced dephasing may wash out the coherence of the system and low temperature atomic vapors would be necessary, or, alternatively, sub-Doppler spectroscopic techniques \cite{PRL.36.1170,PRA.73.062509} could be used. Another envisioned experimental system for DIET is ultra-cold dense atomic clouds \cite{PRA.88.023428}. Such systems are inherently free of inhomogeneous broadening and homogeneous dipole-induced line broadening effects have been observed very recently \cite{Arxiv.2014}. The total number of trapped atoms is however still too limited \cite{Arxiv.2014,PRL.112.113603} to observe the coalescence of two separate resonances. In addition, DIET requires a constant atomic density, and therefore the use of a (quasi) uniform atomic trap \cite{PRL.110.200406,PLT.PRA}.

In the case of multi-level systems, a series of transparency frequencies is expected as a result of Fano-type interferences between closely-spaced energy levels \cite{TBP}. It is therefore anticipated that DIET can be observed in realistic multi-level systems. In this case, each system can be considered as a quantum oscillating dipole with many allowed transitions such as different electronic and/or ro-vibrational levels. Since the prediction and observation of EIT has offered a number of very exciting applications such as slow light \cite{Review_SL} or even stopped light \cite{PRL.84.5094, PRL.86.783, Nature.409.6819, PRL.98.243602, PRA.78.023801}, we can envision similar applications with DIET in the near future. We have also verified that DIET survives with strong incident laser pulses. One may therefore also expect various applications in strong field and attosecond physics, for instance for the generation of high harmonics in dense atomic or molecular gases \cite{Nat.Photon.5.655, Nat.Phys.7.97}.

The authors acknowledge support from the EU (ITN-2010-264951 CORINF) and from the Air Force Office of Scientific Research (Summer Faculty Fellowship 2013).

\bibliographystyle{apsrev4-1}

\end{document}